\documentclass[aps,showpacs,PRB,twocolumn,superscriptaddress]{revtex4-1}
\makeatletter

\newcommand{\Rmnum}[1]{\expandafter\@slowromancap\romannumeral #1@}
\makeatother
\usepackage{amsmath}
\usepackage{graphicx}
\usepackage{subfigure}
\usepackage{color}
\usepackage{amsfonts}
\usepackage{tikz}
\usepackage{esint}
\usepackage{mathrsfs}

\newcommand{\bea}{\begin{eqnarray}}
\newcommand{\eea}{\end{eqnarray}}

\newcommand{\f}{\frac}
\newcommand{\vdimer}{{\vrule height0.2cm width0.05cm depth0pt}}
\newcommand{\hdimer}{{\hrule height0.05cm width0.2cm depth0pt}}

\newcommand{\verdimers}{\hbox{\vdimer \hskip 0.1cm \vdimer}}
\newcommand{\hordimers}{\hbox{\vbox{\hdimer \vskip 0.1cm \hdimer}}}

\begin{document}

\title{Sweeping cluster algorithm for quantum spin systems with strong geometric restrictions}
\author{Zheng Yan}
\email{zhengyan13@fudan.edu.cn}
\affiliation{Department of Physics and State Key Laboratory of Surface Physics, Fudan University, Shanghai 200438, China}
\affiliation{Collaborative Innovation Center of Advanced Microstructures, Nanjing 210093, China}
\author{Yongzheng Wu}
\affiliation{Department of Physics and State Key Laboratory of Surface Physics, Fudan University, Shanghai 200438, China}
\author{Chenrong Liu}
\affiliation{Department of Physics and State Key Laboratory of Surface Physics, Fudan University, Shanghai 200438, China}
\author{Olav F. Sylju{\aa}sen}
\affiliation{Department of Physics, University of Oslo, P. O. Box 1048 Blindern, N-0316 Oslo, Norway}
\author{Jie Lou}
\thanks{Corresponding author}
\email{loujie@fudan.edu.cn}
\affiliation{Department of Physics and State Key Laboratory of Surface Physics, Fudan University, Shanghai 200438, China}
\affiliation{Collaborative Innovation Center of Advanced Microstructures, Nanjing 210093, China}
\author{Yan Chen}
\thanks{Corresponding author}
\email{yanchen99@fudan.edu.cn}
\affiliation{Department of Physics and State Key Laboratory of Surface Physics, Fudan University, Shanghai 200438, China}
\affiliation{Collaborative Innovation Center of Advanced Microstructures, Nanjing 210093, China}

\begin{abstract}
Quantum spin systems with strong geometric restrictions give rise to rich quantum phases such as valence bond solids and spin liquid states. However, the geometric restrictions often hamper the application of sophisticated numerical approaches. Based on the stochastic series expansion method, we develop an efficient and exact quantum Monte Carlo ``sweeping cluster'' algorithm which automatically satisfies the geometrical restrictions. Here we use the quantum dimer model as a benchmark to demonstrate the reliability and power of this algorithm.
Comparing to existing numerical methods, we can obtain higher accuracy results for a wider parameter region and much more substantial system sizes. 
\end{abstract}
\maketitle
\section{introduction}
Frustrated quantum spin systems display rich quantum phases such as valence bond solids\cite{VBS}, resonating valence bond (RVB) states\cite{Anderson1987}, spin ice\cite{Bramwell2001},  and some novel topological states of matter. However, these systems always hamper numerical approaches: exact diagonalization (ED) is limited to finite cluster, quantum Monte Carlo (QMC) has sign problems, and density matrix renormalization group (DMRG)\cite{White1992} works only for (quasi) one-dimensional lattices. So it is challenging to study numerically three-dimensional spin liquids and other nontrivial phases on larger lattices. Nonetheless, such exciting quantum phases are also found in models without geometrical frustration but with strong geometric restrictions. For example, there is no spin liquid in the J-Q model\cite{Lou2009}, but it can be in the quantum dimer model (QDM)\cite{RK1988}. These models are similar, but the QDM has a strong geometric restriction, i.e., there must be only one dimer that belongs to one site. Quantum spin models with geometric restrictions are hard problems even by using sophisticated numerical approaches: it is challenging to do sampling in QMC although it has no sign problem, and it is almost impossible to add blocks in DMRG.

\begin{figure}[htb]
\includegraphics[clip,width=8.6cm]{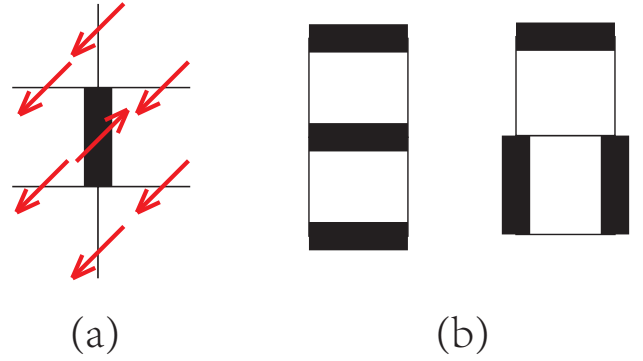}
\caption{(a). The mapping between the link basis and local spin basis. Every link corresponds to a spin site, then up spin indicates that there is a dimer, and down spin indicates that there is a link without dimer. (b). Flip a plaquette (the bottom one) affects the properties of its surrounding plaquettes.}
\label{fig0}
\end{figure}

\begin{figure}[htb]
\includegraphics[clip,width=8.6cm]{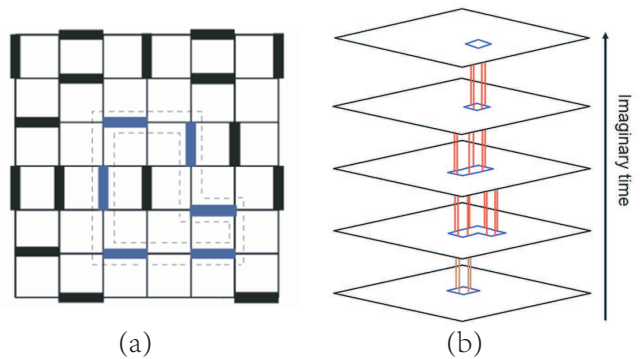}
\caption{(a). Classical loop update of classical dimer models. After flipping all the links enclosed by the dashed lines, you can get a new configuration that obeys the geometric constraints. (b). Schematic diagram of an update for quantum dimer models. Each imaginary time surface is a classical dimer configuration. Red lines are update-lines of world-line QMC. The blue loops are the intersection of all imaginary time update lines and each imaginary time surface which are the same as the classical loop in (a). }
\label{ctoq}
\end{figure}

Usually, the wave function of QDM is written on the link basis(or dimer basis). If the wave function QDM is expressed on the local spin basis which we are familiar with, i.e., every link corresponds to a spin site, then up spin indicates that there is a dimer, and down spin suggests that there is a link without dimer, as depicted in Fig.~\ref{fig0}(a). In terms of local spin basis, geometric restrictions require that six down spins must surround each up spin on a square lattice. This constraint doesn't exist in conventional spin models, like the Heisenberg model or more complicated spin models with 4 spins ring exchange\cite{Melko2005}. On the other hand, in terms of dimer basis, because two plaquettes share a common link, flip a plaquette will affect the properties of its surrounding ones, such as from a flippable plaquette, i.e. plaquette with two parallel dimers, to an unflippable one as shown in Fig.~\ref{fig0}(b). So the update of dimer configuration is not a local effect but a global one. In the classical dimer model, one may use a regular loop update to change dimer configurations as illustrated in Fig.~\ref{ctoq}(a). Connect the thick and thin links into a loop, and flip all the links to get a new configuration which obeys the geometric restrictions. It can be seen that the dimer model is a strongly correlated model. Flipping a link at a location will cause links elsewhere to be flipped to ensure geometric constraints.

The world-line quantum Monte Carlo method maps an n-dimension quantum system into an n+1-D classical system. The +1-D here means the imaginary time dimension. If we want to develop a new QMC method for spin models with constraints such as QDM, its schematic diagram of update must be the same as shown in Fig.~\ref{ctoq}(b): The intersection of all imaginary time update lines and each imaginary time surface must be a classic loop update as the blue loops in this figure. The problem now is how to construct an update method as Fig.~\ref{ctoq}(b) shown following the QMC rules.

In this paper, within the stochastic series expansion (SSE) framework~\cite{Sandvik1991,Sandvik1999}, we develop an efficient QMC algorithm which automatically satisfies the geometric restrictions. In principle, this method works as long as the Hamiltonian does not destroy the geometric constraints. This condition allows us to construct novel quantum states through geometric constraints and study them by QMC. In particular, we use the QDM on square and triangular lattices as examples to elaborate the details of this new algorithm and show that it is efficient by calculating the order parameter on large lattices.

QDMs play an important role as low energy effective descriptions of quantum spin systems~\cite{RK1988,Misguich2003,Poilblanc2010}. The Rokhsar-Kivelson (RK) QDM was first introduced to study quantum spin liquids, and in particular, the physics of the short-range RVB state is probably related to high-Tc cuprates~\cite{Anderson1987,Fazekas1974,Kivelson1987}. Later it was discovered that QDMs also provide particularly simple realizations of topological phases of matter, including a two-dimensional gapped phase with $Z_2$ topological order~\cite{Moessner2001}, and a three-dimensional Coulomb phase described by an emergent $U(1)$ symmetry~\cite{Hermele2004,Huse2003}. Recently, a QDM for the metallic state of the hole-doped cuprates was also proposed to describe the mysterious pseudogap state at low hole density~\cite{Sachdev2015}.

\section{Numerical method}

The QDM Hamiltonian can be written as
\begin{equation}
    H=-\sum_{\rm plaq}\left(\vphantom{\sum}|\verdimers\rangle\langle\hordimers|+\rm{H.c.}\right)
       +V\sum_{\rm plaq}\left(\vphantom{\sum}|\verdimers\rangle\langle\verdimers|+
                                             |\hordimers\rangle\langle\hordimers|\right)
\label{Hamiltonian}
\end{equation}
where the summations are taken over all elementary plaquettes of the lattice. A dimer represents an $SU(2)$ singlet bond between two spins located at its endpoints, and the kinetic term describes a resonance between the two dimerization of a plaquette. This seemingly simple Hamiltonian contains strong geometric constraint which requires every site on the lattice to be covered by one and only one dimer.

The SSE method is a generalization of Handscomb's power series expansion method~\cite{Handscomb} for the isotropic $S=1/2$ Heisenberg ferromagnet and antiferromagnet~\cite{Lyklema1982,Chakravarty1982,DLee1984} to a much wider range of systems.
The starting point of the SSE method is the power series expansion of the partition function in a particular basis $\{|\alpha \rangle\}$.
Generally the $S^z$ basis is chosen for spin systems. For QDMs we choose the dimer basis, and write a dimer basis state as $|\alpha\rangle = |D_1,D_2,\ldots,D_N\rangle$, where $D_i$ takes value 1(0) if there is (not) a dimer on link $i$.

We write the Hamiltonian in terms of plaquette operators $H_p$, $H=-\sum_{p=1}^{N_p}H_p$, where $p$ labels a specific plaquette on the lattice.  The plaquette operators are further decomposed into two operators: $H_p = H_{1,p} + H_{2,p}$, where $H_{1,p}$ is diagonal and $H_{2,p}$ is off-diagonal:
\begin{eqnarray}
H_{1,p} & = &
-V  \left( \vphantom{\sum} | \verdimers \rangle \langle \verdimers |+| \hordimers \rangle \langle \hordimers |\right) + V + C,
\label{hb1} \\
H_{2,p} & = & \left(  \vphantom{\sum} | \verdimers \rangle \langle \hordimers | + | \hordimers \rangle \langle \verdimers | \right).
\label{hb2}
\end{eqnarray}
In this Hamiltonian, we have subtracted a constant $N_p(V+C)$ from Eq.~(\ref{Hamiltonian}), which should be
kept in mind when calculating the energy. We do this because the constant $V+C$ makes all matrix elements of $H_{1,p}$ positive provided $C> {\rm min}(-V,0)$. We will choose $C=1$ here for simplicity.

The powers of $H$ in the series expansion of the partition function $Z$ can be expressed as sums of products of the plaquette operators (\ref{hb1}) and (\ref{hb2}). Such a
product is conveniently referred to by an operator-index sequence: $S_n = [a_1,p_1],[a_2,p_2],\ldots,[a_n,p_n]$, where $a_i \in \lbrace 1,2\rbrace$ corresponds to the type of operator ($1$=diagonal, $2$=off-diagonal) and $p_i \in \lbrace
1,\ldots,N_p\rbrace$ is the plaquette index. It is also convenient to work with a fixed-length operator-index list with $M$ entries and to include the identity operator $[0,0]$ as one of the operator types.

\begin{figure}[htb]
\includegraphics[clip,width=8.6cm]{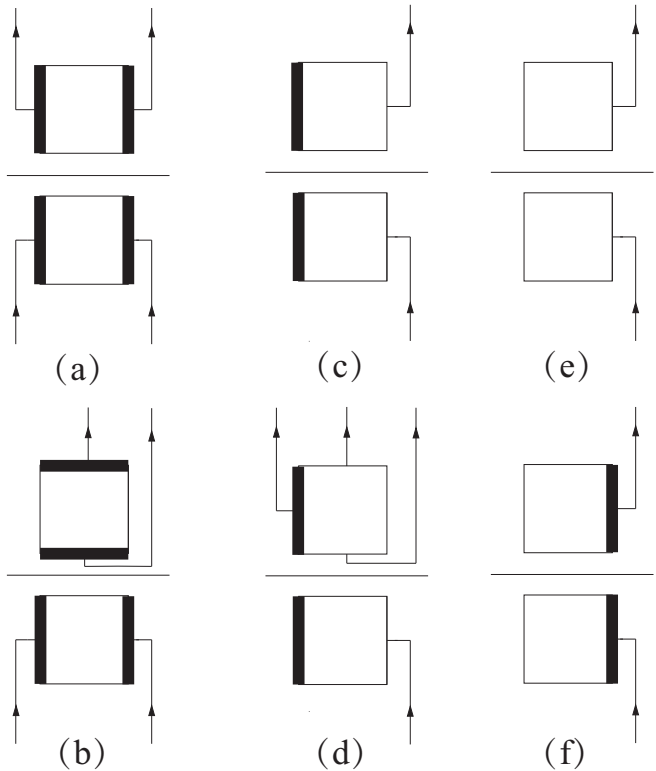}
\caption{Some of the vertices and their update prescriptions. The horizontal bar represents the full plaquette operator $H_p$ and the lines of the squares represent the dimer states (thick and thin lines for dimer 1 or 0) on either side of the operator. Update-lines are shown as lines with an arrow.
(c) and (d) are different updates of the same configuration.}
\label{vertex}
\end{figure}

The expanded partition function takes then the same form as that for the spin models~\cite{Sandvik1991,Sandvik1999},
\begin{equation}
Z = \sum\limits_\alpha \sum_{S_M} {\beta^n(M-n)! \over M!}
    \left \langle \alpha  \left | \prod_{i=1}^M H_{a_i,p_i}
    \right | \alpha \right \rangle ,
\label{zm}
\end{equation}
where $n$ is the number of operators $[a_i,p_i] \not= [0,0]$.
By inserting complete sets of states between all the plaquette operators, the product can be written as a product of the following
non-zero plaquette matrix elements
\begin{eqnarray}
\nonumber
&\langle\verdimers | H_{1,p} | \verdimers \rangle  =
\langle\hordimers | H_{1,p} | \hordimers \rangle  = 1,\\
&\langle\verdimers | H_{2,p} | \hordimers \rangle  =
\langle\hordimers | H_{2,p} | \verdimers \rangle = 1,\\
\nonumber
&\langle {\rm others}   | H_{1,p} | {\rm others}   \rangle  =  1+V,
\label{matrelem}
\end{eqnarray}
the $| {\rm others}\rangle$ here means that plaquette $p$ has 1 or 0 dimer.
Such matrix elements are depicted in Fig.~\ref{vertex} where the plaquette below(above) is the ket(bra).

In the Monte Carlo sampling of the partition function we insert or delete a diagonal operator in the operator-index sequence just like the diagonal update for spins models: We accept the insertion/deletion according to the Metropolis acceptance probabilities,
\begin{eqnarray}
P_{\rm ins} & = &
{N_p\beta \langle \alpha| H_{1,p} | \alpha \rangle \over M-n },
\label{diap1} \\
P_{\rm del} & = &
{M-n+1 \over N_p\beta \langle \alpha| H_{1,p} | \alpha\rangle }.
\label{diap2}
\end{eqnarray}
The presence of $N_p$ in these probabilities reflects the fact that there are $N_p$ random choices for the plaquette $p$ in converting
$[0,0]\to [1,p]$, but only one way to replace $[1,p]\to [0,0]$ when $p$ is given. These diagonal updates are attempted consecutively for all
$1,\ldots,M$, and at the same time the state
$|\alpha \rangle$ is updated when plaquette flipping operators $[2,p]$ are encountered.

Cluster(loop) updates~\cite{Sandvik1999,OFS2002} can accomplish substitutions $[1,p] \leftrightarrow [2,p]$ in the standard scheme applied to spin models. There are several kinds of cluster-update
schemes: operator loop\cite{Sandvik1999}, directed loop\cite{OFS2002}, cluster-like loop\cite{Sandvik2003} and others to solve different models. However, due to the geometric restrictions of the QDM, regular cluster updates cannot be applied. The main result described below is a new kind of cluster update obeying imaginary time order to change operators more efficiently. We call it the ``sweeping cluster'' method. It works as follows.

First, choose a starting operator vertex randomly with flippable plaquettes(FPs) on both sides, either diagonal or off-diagonal. FP means that the plaquette contains two parallel dimers. Next, create a cluster of four update-lines, one for every link of the plaquette, each emanating from the starting vertex in the positive imaginary-time direction. The update-lines serve as guiding lines in the imaginary-time direction on where to change the configuration: The dimer at the end of each update-line is toggled on/off in the proposed new configuration as they sweep simultaneously upwards in imaginary-time. Thus the four initial update-lines rotate the two dimers of the original FP as they go along£¿. The update-lines are extended until they meet another operator vertex from below.
Then, after updating the plaquette beneath on the new operator vertex according to the update-lines, we need to decide how to create or destroy update-lines to update the plaquette above and continue sweeping, see Fig.~\ref{vertex}.

\begin{figure*}[htb]
\includegraphics[clip,width=17cm]{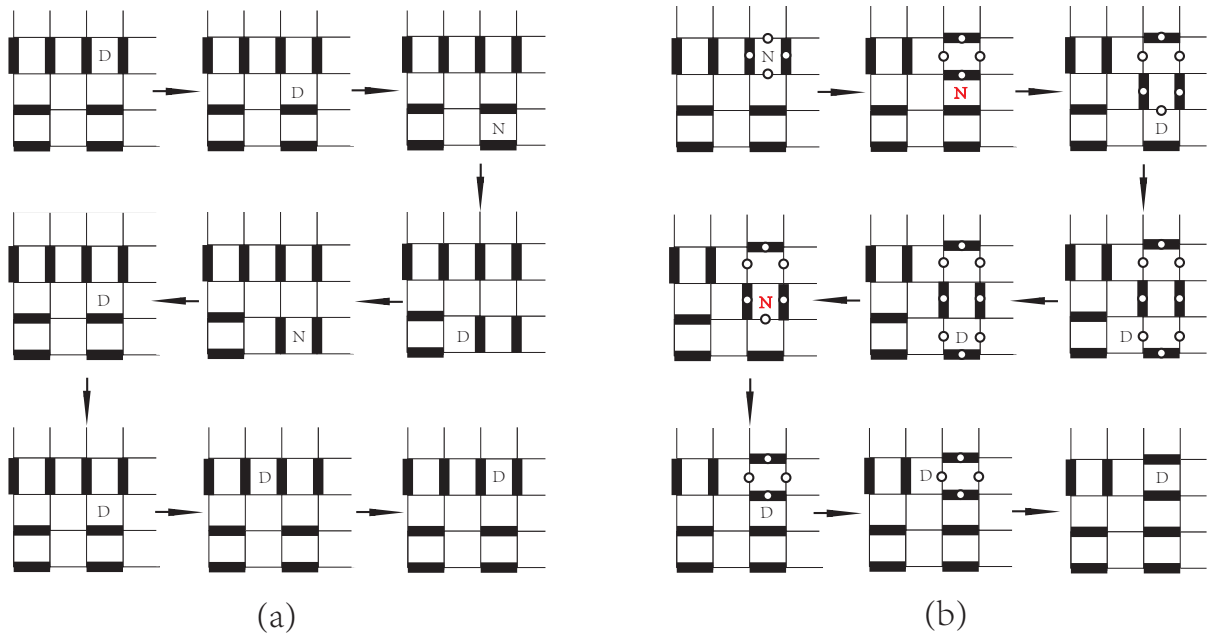}
\caption{Configurations of QDM in imaginary time space. Each picture is a dimer configuration at a certain imaginary time, and the long arrows indicate the increasing imaginary time. The "D" and "N"
means a diagonal and off-diagonal operator. (a) and (b) stand for dimer configurations snapshots before and after the sweeping cluster-update respectively. The red "D" or "N" means half-chance to be a diagonal or off-diagonal operator and we choose the red one randomly.
The circle here means an update-line exist on the link,i.e., the dimer has to be toggled on/off.}
\label{cluster}
\end{figure*}
For this, there are three different processes to consider: (1) The new plaquette beneath is an FP, and the old plaquette above is not an FP. We can then change the plaquette above into an FP in two ways: either the resulting vertex will become diagonal or off-diagonal. We choose between these two possibilities shown in (c) and (d) in Fig.~\ref{vertex} with probability $1/2$. (2) The new plaquette beneath is not an FP. Then the change of the upper plaquette is equivalent to the change of one underneath, as shown in  (a), (b), (e) and (f) in Fig.~\ref{vertex} and the operator should be diagonal. (3) Both the new plaquette beneath and the old plaquette above are FPs. Then there are two choices: the cluster-update ends if the number of total lines is four. If not, the four update-lines continue through the vertex and sweep on. The reason that we keep the operator unaltered in the latter case is to keep a detailed balance regarding its reversed process.

As an example, we draw Fig.~\ref{cluster} where (a) and (b) are the configurations before/after cluster update. Compare the dimer configuration between (a) and (b) at a certain imaginary time, and it returns to loop update in the classical dimer model, i.e., every link passed by loop has to be flipped.

At the end of the sweeping cluster update, when the last four update-lines are deleted, we get a new configuration B with weight $ W_{B} $ to replace the old configuration A with weight $ W_{A} $. To ensure detailed balance, we must invoke a Metropolis accept/reject step\cite{Metropolis} on the whole cluster update with an acceptance probability
\begin{equation}
P_{accept}(A\rightarrow B)  =  \min( {W(B)P_{\rm select}(B\rightarrow A)\over W(A)P_{\rm select}(A\rightarrow B)}, 1 ),
\label{select-accept}
\end{equation}
where $P_{\rm select}(A \rightarrow B)$ is the probability for the sweeping cluster update to change configuration A into B. This step involves both the random choice of starting vertex and the random choices in update type (1). If we denote the number of operator vertices in configuration A with FPs on both sides by $N_{\rm FP}$, and the same amount in configuration B by $N_{\rm FP}+\Delta$, then
\begin{equation}
P_{accept}(A\rightarrow B)  =  \min( \f{N_{\rm FP}}{N_{\rm FP}+\Delta} \left( \f{2}{1+V} \right)^\Delta, 1 ).
\label{select-accept2}
\end{equation}

At low temperature, the first term $\f{N_{\rm FP}}{N_{\rm FP}+\Delta}\approx 1$. At RK point, any new configuration can be accepted. That's because the wave function of the RK point is an equal weight overlap of all configurations.

\section{Results}

\begin{figure}[htb]
\includegraphics[clip,width=8.6cm]{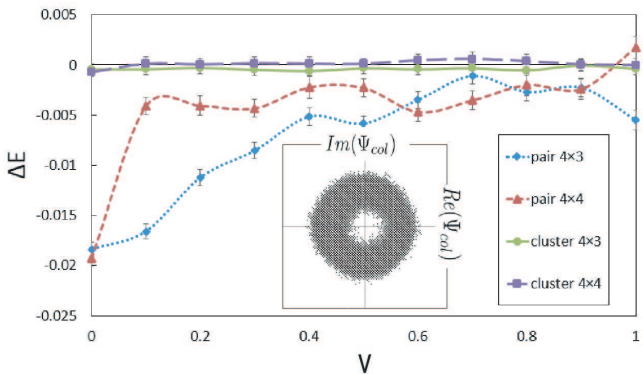}
\caption{Correctness and ergodicity(in a certain winding sector) check: The energy difference between ED and QMC with two distinct updates, pair update and cluster update, on triangle lattice. Inset: Evolution of $\Psi_{col}$ in a complex plane at $V=0.5$ of 16$\times$16 square lattices by serial computing.}
\label{edqmc}
\end{figure}

To demonstrate the potential of our new method, we first show its efficiency. All the following results were obtained under the condition of $T=0.01$. If we want to solve QDM by the old world-line QMC scheme, we can only use ``pair update'' which means flipping two FPs face to face~\cite{Sandvik1991}. This update technique is neither ergodic nor efficient, as can be seen from Fig.~\ref{edqmc}, which shows how much the ``pair update'' and our cluster update deviate from ED for the same number of Monte Carlo steps. Our algorithm matches the ED results much better than the ``pair update'' does. This is because the ``pair update'' only changes a few operators which give long autocorrelation times resulting in statistical errors that are smaller than the real error.

It is also important to check ergodicity (in a certain winding sector) of the method by tracking the movement of the columnar order parameter as defined in Ref.~\cite{Sachdev1989},
\begin{equation}
\begin{split}
\Psi_{col} & = \f{1}{L^2} \sum_{\bf r} \left\{
(-1)^{r_x} [n({\bf r}+{{\bf x}\over 2})-n({\bf r}-{{\bf x}\over 2})]+ \right.\\
                   &\left. i(-1)^{r_y}[n({\bf r}+{{\bf y}\over 2})-n({\bf r}-{{\bf y}\over 2})] \right\},
\label{ficol}
\end{split}
\end{equation}
where {\bf x} and {\bf y} are unit vectors and $L$ is the linear system size. The dimer number operator n({\bf r}+{\bf e}/2) is 1 if the site at {\bf r} and its nearest neighbor at {\bf r}+{\bf e} form a dimer, and zero otherwise. As depicted in the inset of Fig.~\ref{edqmc}, the evolution of $\Psi_{col}$ in a complex plane is circularly distributed even far from the RK point, here we choose $V=0.5$.

\begin{figure}[h]
\includegraphics[clip,width=8cm]{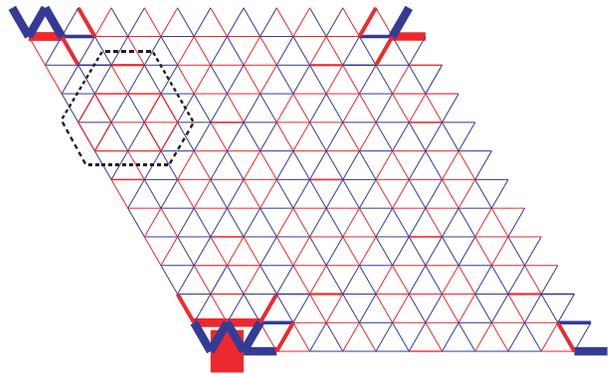}
\caption{The dimer correlation function of QDM on 12$\times$12 triangle lattice at $V=0.5$. We can see a $\sqrt{12} \times \sqrt{12}$ phase clearly. The dashed line helps us to capture the periodic structure.}
\label{triangular}
\end{figure}

On the triangular lattice, there is a novel phase called $\sqrt{12} \times \sqrt{12}$ phase between columnar phase and RVB phase of quantum dimer model~\cite{Ralko2005TR}. By employing our algorithm, we calculate the dimer correlation function as Eq.(\ref{correlationfunction}) of QDM on triangular lattice and obtain this phase as shown in Fig.~\ref{triangular}. Red bonds in this figure corresponds to dimers and blue ones mean no dimer. We can clearly observe the periodic $\sqrt{12} \times \sqrt{12}$ structure unit encircled with the black dashed line. According to the principle of Monte Carlo method, the computational complexity of this algorithm is the same order of magnitude on different lattices, because the units are rotated plaquettes.

\begin{figure}[h]
\includegraphics[clip,width=7.5cm]{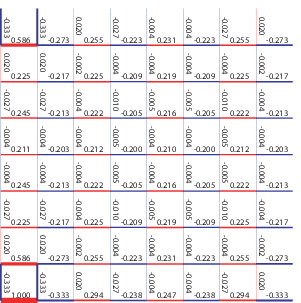}
\caption{The correlation function of QDM on $8\times 8$ square lattice at $V=0$. Red bond (positive number) means dimer strength, blue one (negative number) means no-dimer strength.}
\label{correfunction}
\end{figure}

Hereby we define the dimer correlation function as
\begin{equation}
C_{ij}={{\langle n_in_j \rangle - \langle n_i \rangle \langle n_j \rangle}\over{ \langle n_in_i \rangle- \langle n_i \rangle \langle n_i \rangle}},
\label{correlationfunction}
\end{equation}
$n_i=1(0)$ means link i has a(no) dimer. Furthermore, to verify the accuracy of our algorithm, we also reproduce high precision results for the dimer correlation functions on an $8 \times 8$ square lattice given in Ref.~\cite{Leung1996} which is obtained by ED method. As depicted in Fig.~\ref{correfunction}, we don't label the error bar since our results are within 0.1 percent difference comparing with the ED results.

\begin{figure}[h]
\includegraphics[clip,width=8.4cm]{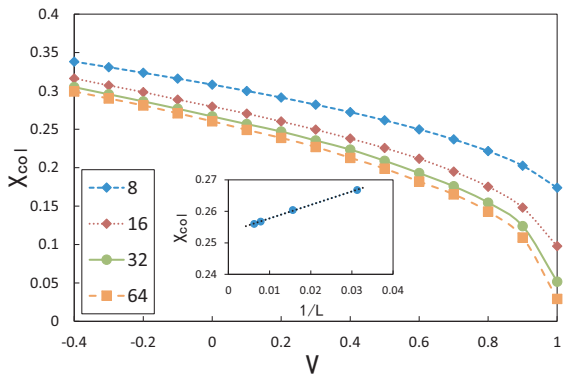}
\caption{The columnar order parameter as a function of $V$ on $L\times L$ square lattices with $L=8,16,32,64$. Inset shows finite size extrapolation at $V=0$ including also data for $L=128$ and $160$.}
\label{order}
\end{figure}

Having established its numerical efficiency and accuracy, we use the method to obtain high-precision results for the QDM. The averaged modulus of the columnar order parameter, $\chi_{col} = \sqrt{\langle | \Psi_{col} |^2 \rangle}$, as a function of $V$ is shown in Fig.~\ref{order} for different lattices sizes. The error bars are smaller than the size of symbol. If long-range columnar order exists, $\chi_{col}$ remains finite as $L \to \infty$. From Fig.~\ref{order} it is seen that $\chi_{col}$ decreases as $L$ gets larger. However, as shown in the inset of Fig.~\ref{order}, an extrapolation carried out for the special value $V=0$, including the results for larger systems up to $L=160$,  indicates that $\chi_{col}$ may converge to a finite value for $L \to \infty$.

Our new method presented here allows the study of QDM on large lattices at finite temperatures. This method is in contrast to zero temperature projector Monte Carlo methods that have only been applied to QDMs of smaller system sizes than used here to keep the statistical errors under control~\cite{OFS2005,Trivedi1989,Baroni1999}. Other quantum cluster algorithm for Ising model with restrictions~\cite{Banerjee2014,Schlittler2015} can be applicable only on specific lattices and certain parameter regions. Another drawback with these methods is that one must ``throw away'' configurations which don't obey the geometric restrictions. This ratio may be as high as $3/4$~\cite{Banerjee2014}.\\

\section{Conclusions and Outlook}
Numerical study of the quantum spin model with strong geometric restrictions is important and notoriously difficult. We have introduced the sweeping cluster SSE method to calculate them. The technique keeps the geometric configuration satisfied by sweeping vertices in imaginary-time order. It is the first finite temperature QMC method for QDMs that samples the dimer space directly, which provides a positive all-around solution to this hard problem. The algorithm is valid and efficient for the whole parameter region of QDMs in principle. It works on any lattice geometries and can be generalized to other models such as quantum loop model~\cite{QLM}. Furthermore, all existing numerical algorithms for quantum dimer model can only do sampling in the same winding sector. We have made progress on realizing the sampling of all winding sectors based on our ¡°sweeping cluster¡± algorithm. Besides, our algorithm is a world-line algorithm. This method provides us with access to the (imaginary-time) dynamic behavior of the quantum dimer model and other spin models with strong geometrical restrictions.

\section{Acknowledgements}
We wish to thank T. K. Lee and Wenan Guo for fruitful discussions. ZY acknowledges the support of Nordic Centre. This work was supported by the State Key Programs of China (Grant Nos. 2017YFA0304204 and 2016YFA0300504), the National Natural Science Foundation of China (Grant Nos. 11625416, and 11474064).

%

\end{document}